\documentclass{PoS}
\newcommand{\hp}{{+}}
\newcommand{\hm}{{-}}
\newcommand{\ber}{\begin{eqnarray}} 
\newcommand{\eer}{\end{eqnarray}}
\newcommand{\cG}{{\cal G}}
\newcommand{\cF}{{\cal F}}

\title{Two-photon exchange corrections in elastic lepton-proton scattering}

\ShortTitle{Two-photon exchange corrections in elastic lepton-proton scattering}

\author{\speaker{Oleksandr Tomalak}
     \\
        Institut f\"ur Kernphysik, Johannes Gutenberg Universit\"at Mainz, 55128 Mainz, Germany\\
        E-mail: \email{tomalak@uni-mainz.de}}

\abstract{We apply a subtracted dispersion relation (DR) formalism with the aim to 
improve predictions for the two-photon exchange (TPE) corrections to elastic electron-proton scattering observables at small momentum transfers. 
We study the formalism on the elastic TPE contribution in comparison with existing data for unpolarized cross sections.

We extend the general formalism of TPE to elastic scattering with massive lepton and perform a numerical estimate of the muon-proton scattering at low momentum transfer in view of the upcoming muon-proton scattering experiment (MUSE). 

We study the influence of the double-virtual Compton scattering (VVCS) subtraction function on the unpolarized lepton-proton scattering cross-section. We show that the resulting TPE correction is negligible in the electron-proton scattering and smaller than planned uncertainties of the MUSE experiment for the subtraction functions evaluated in chiral perturbation theory.}

\FullConference{53rd International Winter Meeting on Nuclear Physics\\
                 26-30 January 2015\\
                 Bormio, Italy}

\begin{document}

Elastic electron-proton scattering in the approximation of the exchange of one-photon provide fundamental information on general properties of the proton. The electric and magnetic form factors of the proton are extracted from the unpolarized elastic $ e p $ cross-section data by the Rosenbluth separation, see Ref. \cite{Bernauer:2013tpr} for the modern extractions. The measurement of the electric over magnetic form factor ratio with polarization transfer from the electron beam to the proton gave strikingly different results \cite{Gayou:2001qd}. TPE contributions have been proposed as a plausible solution to resolve the puzzle. The mysterious discrepancy, often referred to as "the proton radius puzzle", has been observed in the measurement of the proton charge radius from the energy spectrum of the muonic hydrogen compared to the usual hydrogen result  and electron-proton scattering. TPE estimates at small momentum transfer can decrease the uncertainties of the electron-proton scattering data. Their estimates have strong model dependence in the region of the small momentum transfer. For the TPE contribution from the proton intermediate state, that is expected to be the largest TPE effect, we compare the model independent DRs approach \cite{Borisyuk:2008es} with the hadronic model of TPE evaluation \cite{Blunden:2005ew}. In order to avoid the model dependence in the definition of the off-shell photon-proton vertex and with the aim to suppress the contribution of intermediate states with larger invariant masses we use the subtracted DR for one of three independent invariant TPE amplitudes. We make estimate of the leading proton intermediate state TPE correction for the upcoming muon-proton elastic scattering experiment \cite{Gilman:2013eiv}. We study the contribution of the VVCS subtracted function in elastic lepton-proton scattering and reevaluate it in MUSE kinematics.

\section{General formalism of TPE corrections in elastic lepton-proton scattering}

\begin{figure}[h]
\begin{center}
\includegraphics[width=.4\textwidth]{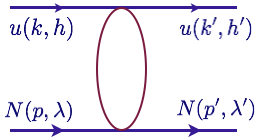}
\end{center}
\caption{Elastic lepton-proton scattering.}
\label{lp_scattering}
\end{figure}

Elastic lepton-proton scattering is completely described by 16 helicity amplitudes $ T_{h' \lambda', h \lambda} $ with arbitrary $h,h',\lambda,\lambda'$ = $\pm $, the sign corresponds to the sign of helicity, see Fig. \ref{lp_scattering}. The QCD and QED discrete symmetries (parity and time-reversal invariance) leave just six independent amplitudes: $ T_1 = T_{\hp \hp, \hp \hp}, ~T_2 = T_{\hp \hm, \hp \hp},  ~T_3 = T_{\hp \hm, \hp \hm} , ~T_4 = T_{\hm \hp, \hp \hp}, ~T_5 = T_{\hm \hm, \hp \hp},  ~T_6 = T_{\hm \hp, \hp \hm} $. 

Helicity amplitudes can be expressed by the sum of six different tensor structures coming together with the invariant amplitudes, that are functions of two kinematic variables. We use the momentum transfer squared $ Q^2 = - ( k - k' )^2 $ and the crossing symmetric variable $ \nu $, that is expressed in terms of the $s$-channel squared energy $ s = ( k + p )^2 $ and the $u$-channel squared energy $ u = ( k - p' )^2 $ as $ \nu = (s-u)/4 $. This variable changes sign with $ s\leftrightarrow u $ channel crossing.

It is convenient to divide the amplitude into a part without the flip of lepton helicity $ T^{\mathrm{nonflip}} $, and the part with lepton helicity-flip $ T^{\mathrm{flip}} $ \cite{Gorchtein:2004ac}. The expression for the helicity amplitude is given by (where the T matrix is defined as $ S = 1 + i T $):
\ber \label{str_ampl} 
T^{\mathrm{nonflip}} & = & \frac{e^2}{Q^2} \bar{u}(k',h') \gamma_\mu u(k,h) \cdot \bar{u}(p',\lambda') \left(\cG_M  \gamma^\mu - \cF_2  \frac{P^{\mu}}{M} + \cF_3  \frac{\gamma . K P^{\mu}}{M^2} \right) u(p,\lambda) , \label{str_ampl1} \\
 T^{\mathrm{flip}} & = &\frac{e^2}{Q^2} \frac{m}{M} \bar{u}(k',h') u(k,h) \cdot \bar{u}(p',\lambda')\left( \cF_4  + \cF_5  \frac{\gamma . K}{M}\right) u(p,\lambda)  \nonumber \\
& + & \frac{e^2}{Q^2} \frac{m}{M} \cF_6  \bar{u}(k',h') \gamma_5 u(k,h) \cdot \bar{u}(p',\lambda') \gamma_5 u(p,\lambda), \label{str_ampl2}
\eer
with $ P = (p+p')/2,~K=(k+k')/2$, the lepton (proton) mass $ m $ ($ M $) and the unit of charge $e$.
 
In the one-photon exchange (OPE) approximation only two amplitudes $ \cG_M $ and $ \cF_2 $ are non zero. The helicity amplitude is expressed in terms of the Dirac $ F_1 $ and Pauli $ F_2 $ FFs as:
\ber \label{OPE_amplitude} 
T  & = &  \frac{e^2}{Q^2} \bar{u}(k',h') \gamma_\mu u(k,h) \cdot \bar{u}(p',\lambda') \Gamma^\mu u(p,\lambda), \\
\Gamma^\mu & = & \gamma^\mu F_1(Q^2) + \frac{i \sigma^{\mu \nu} q_\nu}{2 M} F_2(Q^2).
\eer
It is customary in experimental analysis to work with Sachs magnetic and electric FFs:
\ber  \label{Sachs_ffs}
 G_M  =  F_1 + F_2 , ~~~~~~~
 G_E  = F_1 - \tau F_2,
\eer 
where $ \tau = Q^2 / (4 M^2) $. The exchange of more than one photon gives corrections of order $ O(\alpha) $, with $ \alpha = e^2/(4 \pi) \simeq 1/137 $, to all six amplitudes.

The leading TPE correction to the unpolarized elastic lepton-proton scattering cross-section is given by the interference between the OPE diagram and the sum of box and crossed-box graphs with two photons. We define the TPE correction $ \delta_{2 \gamma} $ through the difference between the cross section with account of exchange of two photons $ \sigma $ and the cross section in the $1 \gamma $-exchange approximation $ \sigma_{1 \gamma} $ by:
\ber
 \sigma = \sigma_{1 \gamma} \left( 1 + \delta_{2 \gamma} \right).
\eer
The leading TPE correction is expressed in terms of the TPE structure amplitudes as:
\ber \label{delta}
\delta_{2\gamma}  & = & \frac{2}{ G_M^2 + \frac{\varepsilon}{\tau} G_E^2} \left\{ G_M \Re \cG_1^{2 \gamma} + \frac{\varepsilon}{\tau} G_E \Re \cG_2^{2 \gamma}  + \frac{ 1 - \varepsilon }{ 1 - \varepsilon_0 } \left( \frac{\varepsilon_0}{\tau}  G_E \Re \cG_4^{2 \gamma}  - G_M \Re \cG_3^{2 \gamma}  \right) \right\} ,
\eer
where we define for convenience the following amplitudes:
\ber
 \cG_1^{2 \gamma}  & = & \cG_M^{2 \gamma}  + \frac{\nu}{M^2} \cF_3^{2 \gamma}  + \frac{m^2}{M^2} \cF_5^{2 \gamma} ,  \\
 \cG_2^{2 \gamma}  & = & \cG_M^{2 \gamma}  - ( 1 + \tau ) \cF_2^{2 \gamma}  + \frac{\nu}{M^2} \cF_3^{2 \gamma} ,  \\
 \cG_3^{2 \gamma}  & = & \frac{m^2}{M^2} \cF_5^{2 \gamma}  + \frac{\nu}{M^2} \cF_3^{2 \gamma} ,   \\
 \cG_4^{2 \gamma}  & = &  \frac{\nu}{M^2}  \cF_4^{2 \gamma}  + \frac{\nu^2}{M^4 (1+\tau)} \cF_5^{2 \gamma} ,
\eer
and use the photon polarization parameter $ \epsilon $:
\ber
\varepsilon = \frac{16 \nu^2 - Q^2 ( Q^2 + 4M^2 )}{16 \nu^2 - Q^2 ( Q^2 + 4M^2 ) + 2 ( Q^2 + 4M^2 )( Q^2 - 2 m^2)},
\eer
The TPE structure amplitudes enter the unpolarized cross-section through their real parts.

\section{Hadronic model versus dispersion relations}

For the region of small energies the TPE correction in elastic lepton-proton scattering can be evaluated by DRs or with assumptions of the photon-proton vertex.

DRs approach is realized for the fixed value of the momentum transfer squared $Q^2$. Structure amplitudes are studied as functions of $ \nu $ variable. The imaginary part of the helicity amplitude $  T_{h' \lambda',h \lambda} $ is obtained from the $S$-matrix unitarity as sum over all possible intermediate states and helicities ($\kappa$): 
\ber
 2 \Im  T_{h' \lambda',h \lambda}  = \sum_{n,\kappa} \prod_{i=1}^n \int \frac{d^3 q_i}{(2 \pi)^3} \frac{1}{2 E_i} T^{*}_{\kappa , h' \lambda'} T_{ \kappa, h \lambda} (2 \pi)^4 \delta^4 (k+k'-\sum_i q_i),
\eer
with $ n $ particles in the intermediate state.
The structure amplitudes are given by the linear combination of the helicity amplitudes.

Assuming analyticity of the structure amplitudes in whole complex $ \nu $ plane, except branch cuts from the particle production threshold to infinity, one can write down DRs for the invariant amplitudes with the fixed value of the squared momentum transfer $ Q^2 $. The invariant amplitudes have definite properties with respect to the change of $ \nu $ variable sign. According to these properties the DRs for the $ ep $ scattering amplitudes $ \cG_M, \cF_2, \cG_1, \cG_2 $ are given by:
\ber
 \Re {\cal{G}}^{2 \gamma} (\nu, Q^2) & = & \frac{2}{\pi} \nu \int \limits^{~~ \infty}_{\nu_{th}} \frac{\Im {\cal{G}}^{2 \gamma}  (\nu', Q^2)}{\nu'^2-\nu^2}  \mathrm{d} \nu',
\eer
and for the amplitude $ \cF_3 $:
\ber
 \Re  {\cal{F}}^{2 \gamma} _{3} (\nu, Q^2) & = & \frac{2}{\pi} \int \limits^{~~ \infty}_{\nu_{th}}  \nu' \frac{\Im  {\cal{F}}^{2 \gamma} _{3}  (\nu', Q^2)}{\nu'^2-\nu^2}  \mathrm{d} \nu' .
 \eer
These DRs reproduce the contribution from the sum of the direct and crossed box graphs.

The proton intermediate state TPE contribution to structure amplitudes can be evaluated also in the hadronic model with the assumption of the on-shell form of the off-shell vertex. For the direct box graph contribution the amplitude is given by:
\ber
 T_{direct}  &  = &   - e^4 \int  \frac{  i \mathrm{d}^4 k_1}{( 2 \pi )^4} \bar{u}(k',h') \gamma^\mu (\hat{k_1}+m) \gamma^\nu u (k,h) \bar{N}(p',\lambda') \Gamma_\mu (\hat{P} + \hat{K} - \hat{k}_1 + M) \Gamma_\nu N (p,\lambda) \nonumber \\
& & \frac{1}{(k_1 - P - K )^2 - M^2} \frac{1}{k_1^2 - m^2} \frac{1}{(k_1 - K - \frac{q}{2} )^2 - \mu^2 } \frac{1}{(k_1 - K + \frac{q}{2} )^2 - \mu^2 }.
\eer
We introduce the finite photon mass $ \mu $ as IR regulator and subtract IR divergences according to the Maximon and Tjon prescription \cite{Maximon:2000hm}, see Sec. \ref{hadronic_muon} for details.

We compare the hadronic model evaluation of the box and crossed box graphs with the evaluation of the amplitudes based on DR formalism for the elastic intermediate state.  For comparison we exploit the dipole form of the electric and magnetic form factors.
We evaluate separately contributions from two Dirac couplings in photon-proton-proton vertex ($ \mathrm{F}_1 \mathrm{F}_1 $), from two Pauli couplings ($ \mathrm{F}_2 \mathrm{F}_2 $) and the mixed case ($\mathrm{F}_1 \mathrm{F}_2 $). We expect different high-energy behaviour of the structure amplitudes for these vertex structures. The imaginary parts of structure amplitudes are the same in both approaches. They are determined by the vertex with on-shell protons. For the real parts of the $ e p $ scattering amplitudes two approaches give the same results for all amplitudes in the case of $ \mathrm{F}_1 \mathrm{F}_1 $ and $ \mathrm{F}_1 \mathrm{F}_2 $ vertex structures and also for the amplitudes $ \cG_1^{2 \gamma}, \cG_2^{2 \gamma}, \cF_2^{2 \gamma} $ in the case of $ \mathrm{F}_2 \mathrm{F}_2 $ vertex structure. The amplitude $ \cF_3^{2 \gamma} $, that was zero in the OPE approximation, differs by a constant. We notice that after one subtraction the DR calculation and the box diagram model results agree.  

\section{Subtracted DR results for unpolarized elastic electron-proton scattering}

In order to avoid the model dependence in the definition of the photon-proton vertex and with the aim to decrease the contribution of resonances and states with higher invariant masses we propose to use the subtracted DR for the amplitude $ \cF_3 $. 

\begin{figure}[h]
\begin{center}
\includegraphics[width=1.\textwidth]{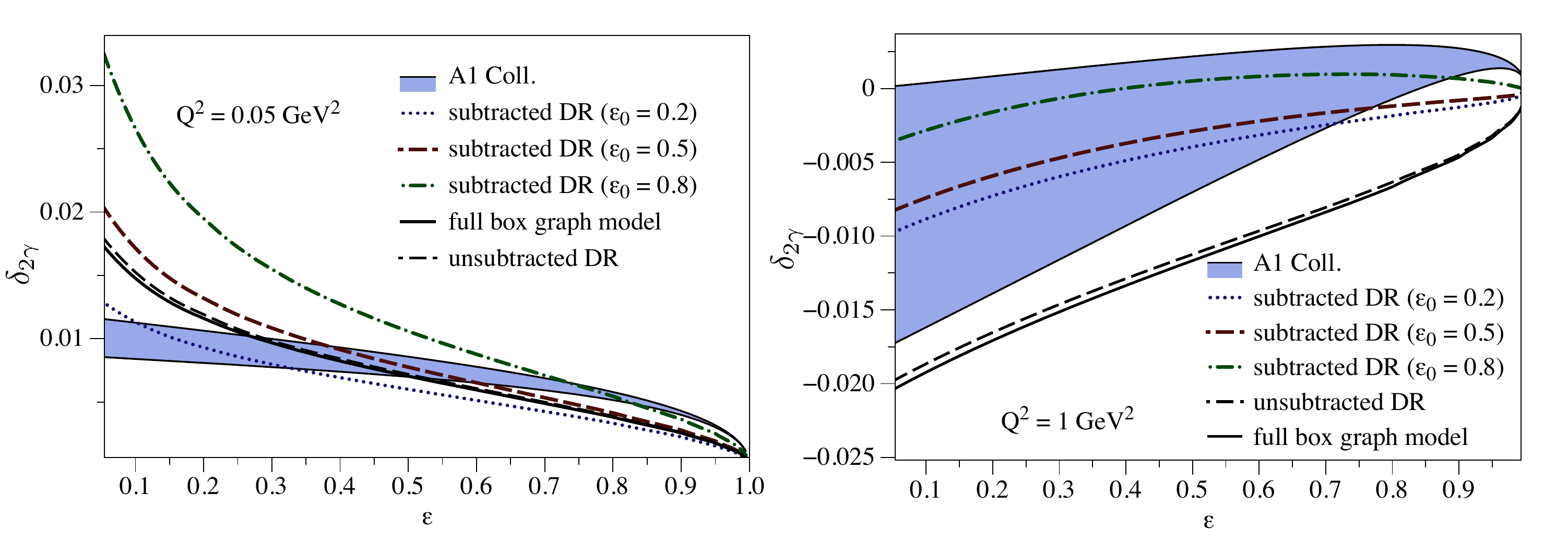}
\end{center}
\caption{Subtracted DR based predictions for the TPE correction for 
$ Q^2 = 0.05 ~\mathrm{GeV}^2 $ (left panel) and 
$ Q^2 = 1 ~\mathrm{GeV}^2 $ (right panel), in comparison with the box diagram model and the unsubtracted DR prediction. Three subtracted DR curves correspond with three choices for the subtraction point~: $ \varepsilon_0 = 0.2,~0.5, ~0.8 $. The blue band is the experimental result from the fit of Ref.~\cite{Bernauer:2013tpr}.}
\label{delta_vs_epsilon_subtracted}
\end{figure}

In the approximation of the elastic intermediate state only the TPE correction to the unpolarized elastic electron-proton scattering cross section can be expressed as the sum of a term evaluated using an unsubtracted DR and a term arising from the $ \mathrm{F}_2 \mathrm{F}_2 $ contribution in the $ {\cal{F}}_3 $ amplitude, which we will evaluate with a subtraction:
\ber
 \delta_{2 \gamma} & = & \delta^{0}_{2 \gamma} + f\left(\nu,Q^2 \right) \Re {\cal{F}}^{F_2F_2}_3, \\
 \delta^0_{2 \gamma} & = &  \frac{2}{G^2_M  + \frac{\varepsilon}{\tau} G^2_E } \left(  G_M \left(\varepsilon -1 \right) \frac{\nu}{M^2} \Re {\cal{F}}^{F_1F_1+F_1F_2}_3  + G_M  \Re  {\cal{G}}_1 + \frac{\varepsilon}{\tau} G_E  \Re  {\cal{G}}_2  \right), \\
f(\nu,Q^2) & = & \frac{2 {G_M} (\varepsilon -1 ) }{G^2_M  + \frac{\varepsilon}{\tau} G^2_E }  \frac{\nu}{M^2}.
\eer

The predictions for the elastic electron-proton scattering cross section correction can be made with one subtraction point at $ \nu = \nu_0 $, the correction is expressed as:
\ber
 \delta_{2 \gamma}(\nu,Q^2) & = &   f(\nu,Q^2) \left[ \Re {\cal{F}}^{F_2F_2}_3 (\nu,Q^2) - \Re {\cal{F}}^{F_2F_2}_3 (\nu_0,Q^2) \right]  \nonumber \\
 & + & \delta^{0}_{2 \gamma}(\nu,Q^2)  + f(\nu,Q^2) \, \Re {\cal{F}}^{F_2F_2}_3 (\nu_0,Q^2)
 \label{delta_prediction}, 
 \eer
 where we can express the value of the subtraction function $ \Re {\cal{F}}^{F_2F_2}_3 (\nu_0,Q^2) $ through the experimental input $  \delta_{2 \gamma}(\nu_0,Q^2) $, that we obtain from the fit of the experimental data \cite{Bernauer:2013tpr}, as:
 \ber \label{subtraction_term}
 \Re {\cal{F}}^{F_2F_2}_3 (\nu_0,Q^2) = \frac{\delta_{2 \gamma}(\nu_0,Q^2) - \delta^{0}_{2 \gamma}(\nu_0,Q^2) }{f(\nu_0,Q^2)}. 
 \eer
 
We show the realization of this formalism \cite{Tomalak:2014sva} for two different values of the momentum transfer in Fig. \ref{delta_vs_epsilon_subtracted}. We take three different subtraction points. The space between curves describes the error of our analysis coming from the inelastic intermediate states, the assumptions of the fit details and the assumption of the dipole form of electric and magnetic proton form factors.

\section{Hadronic model estimates of the elastic TPE correction in MUSE experiment}
\label{hadronic_muon}

We make estimates of the TPE corrections in MUSE experiment coming from the elastic intermediate state in the hadronic model described above. We subtract the IR divergencies according to the Maximon and Tjon prescription \cite{Maximon:2000hm}. The QED TPE amplitude $ \cG_M $ for the scattering of two point-like charges (i.e., $ \mathrm{F}_1 \mathrm{F}_1 $ contribution with $ F_1 ( Q^2 ) = 1 $) has the following IR divergent term:
\ber
 \cG^{IR,0}_M & = &    \frac{s - M^2 - m^2}{\sqrt{\Sigma_s}}   \left(\ln \left(\frac{\sqrt{\Sigma_s}-s+(m+M)^2}{\sqrt{\Sigma_s}+s-(m+M)^2}\right)+i \pi \right) \frac{\alpha}{\pi}  \ln \left(- \frac{t}{\mu^2}\right) \nonumber \\
 & - & \frac{u - M^2 - m^2}{\sqrt{\Sigma_u}}  \ln \left(\frac{\sqrt{\Sigma_u}-u+(m+M)^2}{ - \sqrt{\Sigma_u} -u + (m+M)^2}\right) \frac{\alpha}{\pi}  \ln \left(- \frac{t}{\mu^2}\right), 
\eer
with $  \Sigma_s \equiv (s-(m+M)^2)(s-(m-M)^2) $. The IR divergent contribution to the $ \cG_M^{2 \gamma} $ amplitude is given by $ F_1(Q^2) \cG^{IR,0}_M  $ for the $ \mathrm{F}_1 \mathrm{F}_1 $ vertex structure. The IR divergent contribution to $ \cG_M^{2 \gamma} $ and $ \cF_2^{2 \gamma} $ is given by $  F_2(Q^2) \cG^{IR,0}_M $ for the $ \mathrm{F}_1 \mathrm{F}_2 $ vertex structure. The other amplitudes are IR finite in case of the $ \mathrm{F}_1 \mathrm{F}_1 $ and $ \mathrm{F}_1 \mathrm{F}_2 $ vertex structures. The $ \mathrm{F}_2 \mathrm{F}_2 $ vertex structure is IR finite.

\begin{figure}[h]
\begin{center}
\includegraphics[width=1.\textwidth]{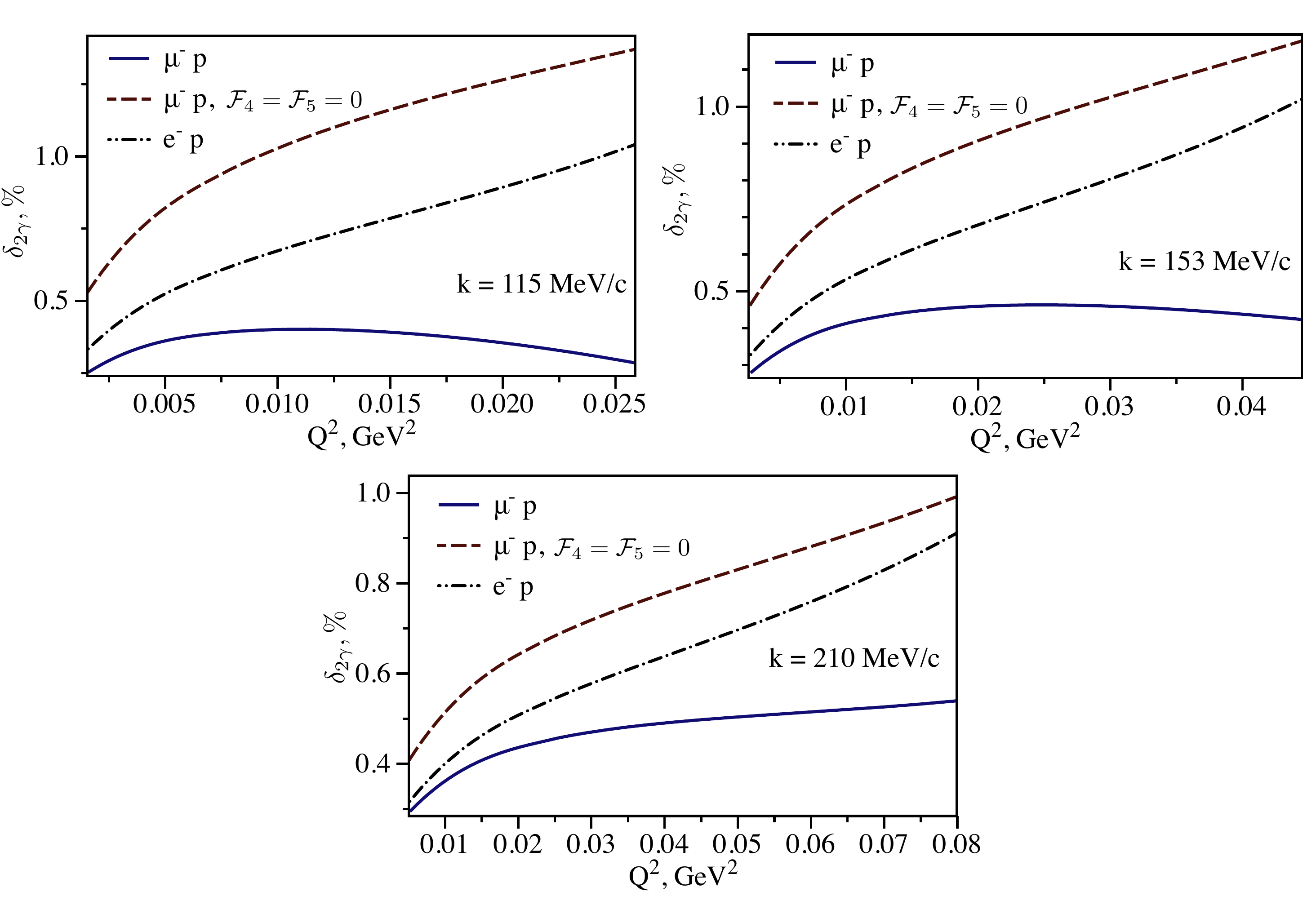}
\end{center}
\caption{The TPE correction $ \delta_{2 \gamma} $ is shown as function of the momentum transfer for three nominal beam momentas of the MUSE experiment. We show the full correction for muon-proton (blue solid line) and electron-proton (black dash-dotted line) scattering together with the muon-proton scattering correction without the flip of muon helicity (red dashed line).}
\label{tpe_electron_muon}
\end{figure}

We show our results \cite{Tomalak:2014dja} in Fig. \ref{tpe_electron_muon}  for the nominal beam momentas of MUSE experiment. For the small momentum transfer the TPE correction has the same order of magnitude as the expected uncertainty of the MUSE experiment.

\section{Subtraction function contribution to elastic unpolarized scattering}

The leading term in the momentum transfer expansion of the TPE contribution from the inelastic intermediate states can be evaluated in the approximation of the {\it forward} VVCS tensor as the lower blob of the TPE graph. The covariant tensor structure of the unpolarized part of the {\it forward} VVCS can be written as:
\ber
\label{vvcs_agreement_with_elastic}
M^{\mu \nu}(\tilde{\nu},\tilde{Q}^2)& = &
 - \left( -g^{\mu\nu}+\frac{\tilde{q}^{\mu} \tilde{q}^{\nu}}{\tilde{q}^2}\right)
T_1(\tilde{\nu}, \tilde{Q}^2 ) - \nonumber \\
& & \frac{1}{M^2} \left(P^{\mu}+\frac{ M \tilde{\nu} }{\tilde{Q}^2 }\,\tilde{q}^{\mu}\right) \left(P^{\nu}-\frac{ M \tilde{\nu} }{\tilde{Q}^2  }\, \tilde{q}^{\nu} \right) T_2 (\tilde{\nu}, \tilde{Q}^2).
\eer%
The unpolarized forward VVCS process is described by two invariant amplitudes $T_1, T_2$, which are functions of $ \tilde{Q}^2 \equiv -\tilde{q}^2$ and $ \tilde{\nu} \equiv P \cdot \tilde{q} / M$. 

The amplitude $ T_2 $ can be completely evaluated within DRs from the proton inelastic structure function, while the amplitude $ T_1 $ is extracted up to the subtraction function $ T_1 (0,\tilde{Q}^2) $ \cite{Carlson:2011zd}, that can be evaluated in chiral perturbation theory \cite{Birse:2012eb,Alarcon:2013cba} up to few hundreds $ \mathrm{MeV} $. In Ref. \cite{Miller:2012ne} there was an attempt to solve the radius puzzle with the enhanced $ T_1(0, \tilde{Q}^2) $ function at large momentum transfer region. It is interesting to study its contribution to the unpolarized scattering cross section. The description of the VVCS process by two amplitudes is valid only for the energies much smaller than the hadronic scale, consequently one can not provide the reliable prediction for the contribution of such subtraction functions. We assume the VVCS description by Eq. (\ref{vvcs_agreement_with_elastic}) to be valid.

It turns out, that the correction is proportional to the squared lepton mass and therefore can be neglected in the measurement of the unpolarized electron-proton scattering cross-section. The subtraction in VVCS amplitude corresponds with the subtraction in the $ {\cal{F}}_4 $ invariant amplitude of the lepton-proton elastic scattering. For the region of small $ Q^2 $ the TPE contribution from the enhanced at large momentum transfer $ T_1(0, Q^2) $ can be approximated as:
\ber
 \delta^{subt}_{2 \gamma,0} = - Q^2 \frac{3 m^2}{2 E} \int \limits^{~~\infty}_0  \frac{T_1 (0, \tilde{Q}^2)}{\pi} \frac{\mathrm{d} \tilde{Q}^2}{\left( \tilde{Q}^2 \right)^2} ,
\eer
with the lepton momentum $ p $ and the lepton energy $ E $. Substituting the enhanced at large momentum transfer subtraction function we reproduce the result of the Ref. \cite{Miller:2012ne} .

\begin{figure}[h]
\begin{center}
\includegraphics[width=.65\textwidth]{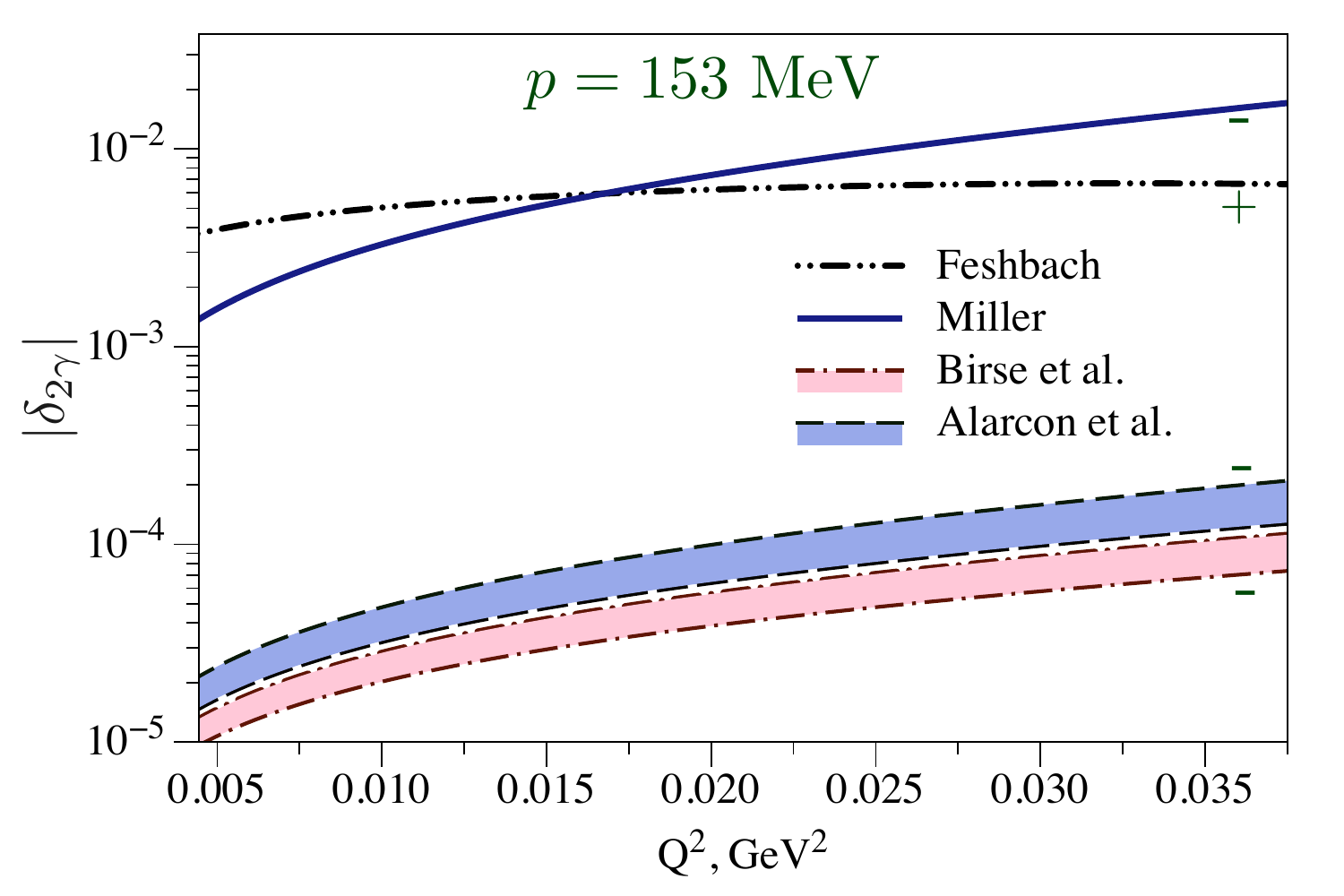}
\end{center}
\caption{The forward Compton scattering subtraction function $ T_1 (0,Q^2) $ contribution to TPE correction in elastic muon-proton scattering for the beam momenta $ p = 153 ~\mathrm{MeV} $. The TPE correction from the different subtraction functions of Refs. \cite{Birse:2012eb,Alarcon:2013cba,Miller:2012ne} is shown in comparison with the leading elastic contribution given by the Feshbach term. The sign corresponds to the $ \mu^- p $ scattering.}
\label{delta_TPE_Q2_E_210}
\end{figure}

In Fig. \ref{delta_TPE_Q2_E_210} we show the contribution from the subtraction functions of Refs. \cite{Birse:2012eb,Alarcon:2013cba,Miller:2012ne} for the typical kinematics of the MUSE experiment. We estimate uncertainties of the baryon chiral perturbation theory \cite{Alarcon:2013cba} varying the upper integration limit between 0.5 $\mathrm{GeV^2}$ and 10 $\mathrm{GeV^2}$. The effect coming from the enhanced at large $ \tilde{Q}^2 $ subtraction function is of the same order as the expected uncertainty of the MUSE experiment or slightly below it, while the more realistic theoretical predictions are few orders of magnitude smaller.

\section*{Acknowledgements}
I thank Marc Vanderhaeghen for the supervision during this work. This work was supported by the Deutsche Forschungsgemeinschaft DFG in part through the Collaborative Research Center [The Low-Energy Frontier of the Standard Model (SFB 1044)], in part through the Graduate School [Symmetry Breaking in Fundamental Interactions (DFG/GRK 1581)].

\end{document}